\documentclass[twoside,twocolumn,9pt]{article}
\usepackage{extsizes}
\usepackage[super,sort&compress,comma]{natbib} 
\usepackage[version=3]{mhchem}
\usepackage[left=1.5cm, right=1.5cm, top=1.785cm, bottom=2.0cm]{geometry}
\usepackage{balance}
\usepackage{mathptmx}
\usepackage{sectsty}
\usepackage{graphicx} 
\usepackage{lastpage}
\usepackage[format=plain,justification=justified,singlelinecheck=false,font={stretch=1.125,small,sf},labelfont=bf,labelsep=space]{caption}
\usepackage{float}
\usepackage{fancyhdr}
\usepackage{fnpos}
\usepackage[english]{babel}
\addto{\captionsenglish}{%
}
\usepackage{array}
\usepackage{droidsans}
\usepackage{charter}
\usepackage[T1]{fontenc}
\usepackage[usenames,dvipsnames]{xcolor}
\usepackage{setspace}
\usepackage[compact]{titlesec}
\usepackage{hyperref}

\usepackage{epstopdf}

\definecolor{cream}{RGB}{222,217,201}

\begin{document}

\pagestyle{fancy}
\thispagestyle{plain}
\fancypagestyle{plain}{
\renewcommand{\headrulewidth}{0pt}
}

\makeFNbottom
\makeatletter
\renewcommand\LARGE{\@setfontsize\LARGE{15pt}{17}}
\renewcommand\Large{\@setfontsize\Large{12pt}{14}}
\renewcommand\large{\@setfontsize\large{10pt}{12}}
\renewcommand\footnotesize{\@setfontsize\footnotesize{7pt}{10}}
\makeatother

\renewcommand{\thefootnote}{\fnsymbol{footnote}}
\renewcommand\footnoterule{\vspace*{1pt}%
\color{cream}\hrule width 3.5in height 0.4pt \color{black}\vspace*{5pt}} 
\setcounter{secnumdepth}{5}

\makeatletter 
\renewcommand\@biblabel[1]{#1}            
\renewcommand\@makefntext[1]%
{\noindent\makebox[0pt][r]{\@thefnmark\,}#1}
\makeatother 
\renewcommand{\figurename}{\small{Fig.}~}
\sectionfont{\sffamily\Large}
\subsectionfont{\normalsize}
\subsubsectionfont{\bf}
\setstretch{1.125} 
\setlength{\skip\footins}{0.8cm}
\setlength{\footnotesep}{0.25cm}
\setlength{\jot}{10pt}
\titlespacing*{\section}{0pt}{4pt}{4pt}
\titlespacing*{\subsection}{0pt}{15pt}{1pt}

\fancyfoot{}
\fancyfoot[LO,RE]{\vspace{-7.1pt}\includegraphics[height=9pt]{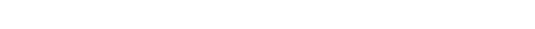}}
\fancyfoot[CO]{\vspace{-7.1pt}\hspace{13.2cm}\includegraphics{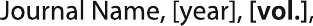}}
\fancyfoot[CE]{\vspace{-7.2pt}\hspace{-14.2cm}\includegraphics{head_foot/RF}}
\fancyfoot[RO]{\footnotesize{\sffamily{1--\pageref{LastPage} ~\textbar  \hspace{2pt}\thepage}}}
\fancyfoot[LE]{\footnotesize{\sffamily{\thepage~\textbar\hspace{3.45cm} 1--\pageref{LastPage}}}}
\fancyhead{}
\renewcommand{\headrulewidth}{0pt} 
\renewcommand{\footrulewidth}{0pt}
\setlength{\arrayrulewidth}{1pt}
\setlength{\columnsep}{6.5mm}
\setlength\bibsep{1pt}

\makeatletter 
\newlength{\figrulesep} 
\setlength{\figrulesep}{0.5\textfloatsep} 

\newcommand{\topfigrule}{\vspace*{-1pt}%
\noindent{\color{cream}\rule[-\figrulesep]{\columnwidth}{1.5pt}} }

\newcommand{\botfigrule}{\vspace*{-2pt}%
\noindent{\color{cream}\rule[\figrulesep]{\columnwidth}{1.5pt}} }

\newcommand{\dblfigrule}{\vspace*{-1pt}%
\noindent{\color{cream}\rule[-\figrulesep]{\textwidth}{1.5pt}} }

\makeatother

\twocolumn[
  \begin{@twocolumnfalse}
{\includegraphics[height=30pt]{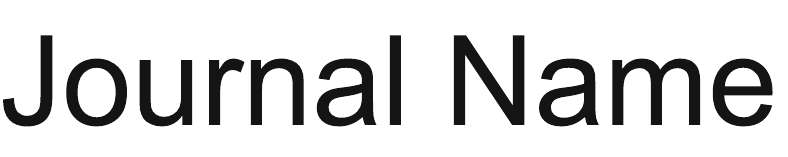}\hfill\raisebox{0pt}[0pt][0pt]{\includegraphics[height=55pt]{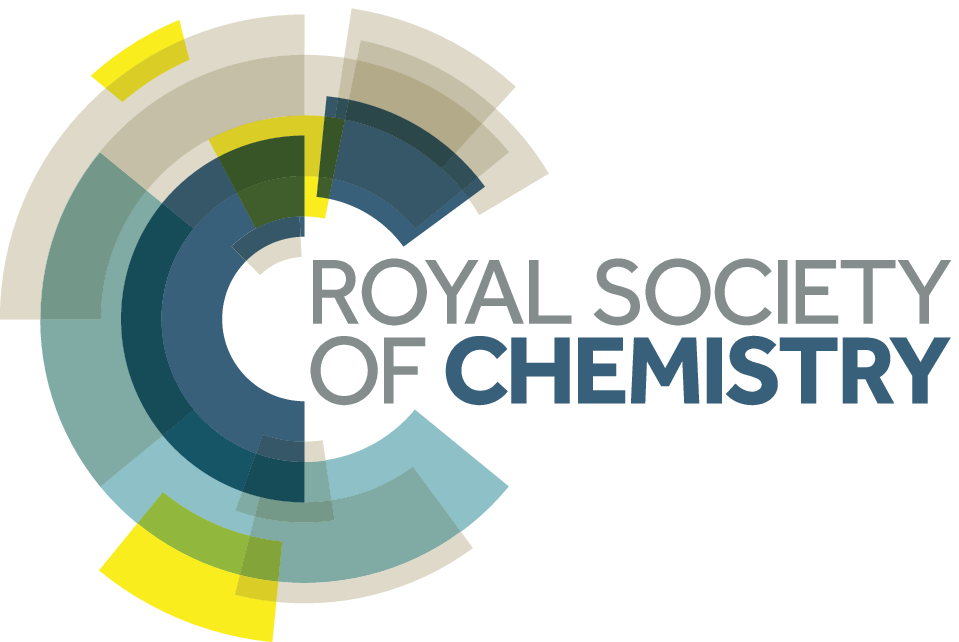}}\\[1ex]
\includegraphics[width=18.5cm]{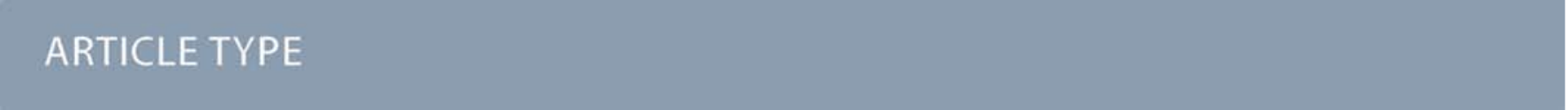}}\par
\vspace{1em}
\sffamily
\begin{tabular}{m{4.5cm} p{13.5cm} }

\includegraphics{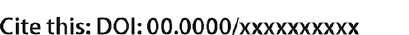} & \noindent\LARGE{\textbf{Realization of the electric-field driven ``one-material''-based magnetic tunnel junction using van der Waals antiferromagnetic MnPX$_3$ (X: S, Se)$^\dag$}} \\
\vspace{0.3cm} & \vspace{0.3cm} \\

 & \noindent\large{Yichen Jin,\textit{$^{a}$} Mouhui Yan,\textit{$^{a}$} Yuriy Dedkov,$^{\ast}$\textit{$^{a,b}$} and Elena Voloshina$^{\ast}$\textit{$^{a,b,c}$}} \\

\includegraphics{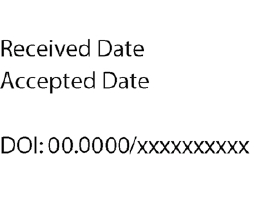} & \noindent\normalsize{Presently a lot of efforts are devoted to the investigation of new two-dimensional magnetic materials, which are considered as promising for the realization of the future electronics and spintronics devices. However, the utilization of these materials in different junctions requires complicated processing that in many cases leads to unwanted parasitic effects influencing the performance of the junctions. Here, we propose the new elegant approach for the realization of the ``one-material''-based magnetic tunnel junction. The several layers of 2D van der Waals MnPX$_3$ (X: S, Se), which is insulating antiferromagnet in its ground state, are used and the effect of the applied external electric filed leads to the half-metallic ferromagnetic states for the outermost layers of the MnPX$_3$ stack. The rich states diagram of such magnetic tunnel junction permits to precisely control its tunneling conductivity. The realized ``one-material''-based magnetic tunnel junction allows to avoid all effects connected with the lattice mismatches and carriers scattering effects at the materials interfaces, giving high perspectives for the application of such systems in electronics and spintronics.} \\

\end{tabular}

 \end{@twocolumnfalse} \vspace{0.6cm}

  ]

\renewcommand*\rmdefault{bch}\normalfont\upshape
\rmfamily
\section*{}
\vspace{-1cm}


\footnotetext{\textit{$^{a}$~Department of Physics, Shanghai University, 99 Shangda Road, 200444 Shanghai, P. R. China. 
E-mail: yuriy.dedkov@icloud.com; elena.voloshina@icloud.com}}
\footnotetext{\textit{$^{b}$~Centre of Excellence ENSEMBLE3 Sp.z\,o.\,o., Wolczynska Str. 133, 01-919 Warsaw, Poland. }}
\footnotetext{\textit{$^{c}$~Institute of Chemistry and Biochemistry, Freie Universit\"at Berlin, Arnimallee 22, Berlin, Germany. }}

\footnotetext{\dag~Electronic Supplementary Information (ESI) is available free of charge. See DOI: 10.1039/D1TC05922J.}



\section{Introduction}

Spintronics is an emerging form of electronics that uses the spin degree of freedom for electrons as an information carrier to achieve data storage and logical operations~\cite{Bader.2010,Zutic.2004}. Compared to conventional microelectronic devices based on charge, spintronics devices require less energy to switch a spin state, which can result in faster operation speed and lower energy consumption. In real applications, spintronics faces a number of challenges, such as generation of fully spin-polarised carriers and injection of spin into devices, long distance spin transport, and manipulation and detection of carriers’ spin orientation. In solving these problems, new concepts and materials for spintronics were proposed, such as the use of semimetals~\cite{Araki.2018}, spin gapless semiconductors~\cite{Wang.2008} and bipolar magnetic semiconductors~\cite{Li.2012}. Within this trend, two-dimensional (2D) van der Waals (vdW) materials are at the frontier of material research since the isolation of graphene (gr)~\cite{Falko.2007,Trauzettel.2007}.

One of the exciting examples on the application of 2D materials, and particularly of graphene and its isostructural counterpart hexagonal BN (h-BN), in spintronics is the recent proposal to use them as an ideal spin-filtering materials in the gr/Ni and gr/Co junctions, which efficiency rapidly reaches $100$\% with increasing the number of graphene layers~\cite{Karpan.2007,Karpan.2008,Karpan.2011}. Here, the unique combination of the electronic structures of graphene and Ni(111) or Co(0001) surfaces permits the transport across the gr/ferromagnet (FM) interface of only electrons with one kind of spin for the states around the $K$ point of the hexagonal Brillouin zone. However, the mixing of the electronic states at the gr/FM interface~\cite{Dedkov.2010,Usachov.2015} leads to the dramatic changes in the spin-transport properties for such interfaces, that can be overcome by using noble metals or h-BN monolayers at the gr/FM interface~\cite{Karpan.2007,Karpan.2008,Karpan.2011} leading to the complexity in the system preparation. Another example is the attempt to use h-BN (and other 2D materials) as insulating layers in the FM/insulator/FM magnetic tunnel junctions (MTJs), which tunnelling magneto-resistance (TMR) depends on the relative magnetization orientations of two FM layers and in the ideal case should reach $100$\% (the so-called ``pessimistic'' formula for TMR)~\cite{Faleev.2015,Piquemal-Banci.2016,Piquemal-Banci.2017}. However, the maximal value of $6$\% was obtained for the Fe/$1$\,ML\,h-BN/Co MTJ~\cite{Piquemal-Banci.2016} which can be connected with the hybridization of the electronic states at the h-BN/FM interface~\cite{Faleev.2015,Tonkikh.2016}. All these difficulties including also the fundamental problems, like, e.\,g., the mismatch in the conductivity of the metal and barrier material (gr), stimulate further studies of the low-dimensional materials and systems which can be used in spintronics applications. For example, further experiments on the 2D materials as tunneling barriers lead to the recent application of the stacking-dependent ferroelectricity in bilayer h-BN in tunnel junctions~\cite{Tsymbal.2006,Yasuda.2021}.  

The general expansion to the two- and tri-atomic 2D materials allows to overcome some disadvantages of graphene and h-BN, like the absence of the energy gap in the carriers' spectrum of graphene and technical difficulties in preparation of high-quality multilayers of graphene and h-BN, however, bringing other challenges, like a scalability. Here, the big advantage in the use of single- or multi-layers of these complex van der Waals materials is their intrinsic purity as well as the ideal interfaces between layers, that reduces unwanted scattering of tunnelling electrons in possible electronic applications. Besides the widely studied well-known transition metal dichalcogenides (TMDs), like MoS$_2$ or WSe$_2$~\cite{Manzeli.2017,Chowdhury.2020}, in recent years the increased attention has been paid to a new class of layered van der Waals materials, transition-metal phosphorus trichalcogenides (TMTs) MPX$_3$ with M: transition metal and X: S, Se~\cite{Wang.2018flk,Samal.2020}. The TMT materials can be easily synthesized using chemical vapour transport method with a size of high-quality bulk crystals up to $1\,\mathrm{cm}^2$ and then the 3D MPX$_3$ crystals can be easily exfoliated into 2D honeycomb multilayers or monolayers~\cite{Du.2018,Jenjeti.2018,Dedkov.2020,Yan.2021ni}. According to the recent studies, most of MPX$_3$ 3D bulk and 2D monolayers are semiconductors with a band gap ranging between $1.3$\,eV and $3.5$\,eV and have antiferromagnetic (AFM) arrangement of magnetic moments of transition metal ions~\cite{Chittari.2016,Yang.2020uc,Yang.2020,Xu.2021fg}. At that, the magnetic phase transition for the MPX$_3$ materials to the FM state can be realised through applying stress~\cite{Chittari.2016} or modulating the carrier concentration via alloying or using field effect~\cite{Yang.2020,Li.2014}. 

\begin{figure}
\includegraphics[width=0.48\textwidth]{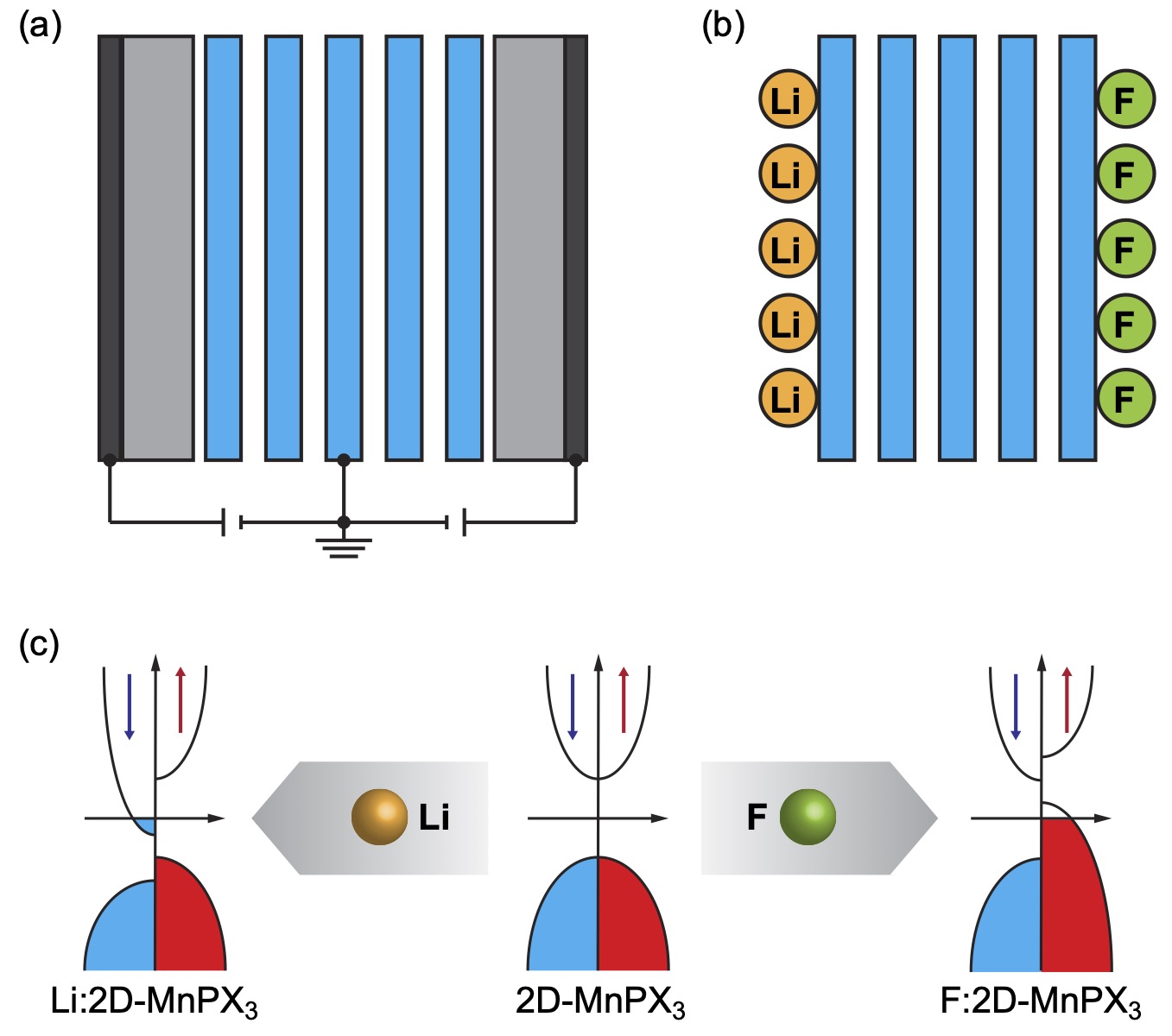}
\caption{\label{fig:scheme} Schematic representation of the electric-field driven ``one-material''-based tunnel junction. (a) Several layers of MnPX$_3$ are placed in the field-effect structure. The electron/hole doping is induced in the left and right outermost layers of MnPX$_3$ using applied bias voltages. (b) Two outermost MnPX$_3$ layers of the TMT stack are covered with Li and F, which mimic the external electric field. The mechanism of induction of the HMF state in 2D-MPX$_3$ with the sign of spin polarization depending on the kind of adsorbate is shown schematically in (c).}
\end{figure}

Here, we demonstrate that many present-day drawbacks in the realization of the spin-filtering devices and MTJs on the basis of 2D materials can be overcome by using TMT MnPX$_3$ stack. In such ``one-material''-based tunnel junction (Fig.~\ref{fig:scheme}a), several layers of bulk-like MnPX$_3$ are sandwiched in the field-effect transistor structure with a bias voltage of the same (or different) polarity applied to the metallic electrodes separated from the outmost MnPX$_3$ layer by a thin insulating layer. As was mentioned, the single 2D MnPX$_3$ layer can be converted from the AFM to HMF states using electric field and the central layer(s) of MnPX$_3$connected to the zero potential via ohmic contact. In the proposed structure, we take the advantage of the relatively large van der Waals gap (around $3.3$\,\AA) and very weak electronic interaction between single MnPX$_3$ layers. In this case, the left/right outermost MnPX$_3$ layers can be converted into the HMF state using external electric field or adsorbates, and at the same time without changing the magnetic (AFM) and insulating state of the inner layers. Such configuration has the TMR value of $100$\%. As it is shown, the different induced electron/hole doping levels of the left/right outermost MnPX$_3$ layers can lead to a variable Curie temperature ($T_C$) for these layers, that gives, together with a N\'eel temperature ($T_N$) for the internal AFM layers, a possibility to realize different magnetic configurations just in ``one-material''-based tunnel junction.

\section{Computational details}

The spin-polarized DFT calculations were carried out with the Vienna \textit{ab initio} simulation package (VASP)~\cite{Kresse.1996,Kresse.1994,Kresse.1993}, employing the generalised gradient approximation (GGA) functional of Perdew, Burke and Ernzerhof (PBE)~\cite{Perdew.1996}. The ion-cores were described by projector-augmented-wave (PAW) potentials~\cite{Bloechl.1994} and the valence electrons [Mn ($3p$, $3d$, $4s$), P ($3s$, $3p$), S ($3s$, $3p$), Se ($4s$, $4p$), Li  ($1s$, $2s$), and F ($2s$, $2p$)] were described by plane waves associated to kinetic energies of up to $500$\,eV. Brillouin-zone integration was performed on $\Gamma$-centred symmetry with Monkhorst-Pack meshes by Gaussian smearing with $\sigma$ = 0.05 eV except for density of states (DOS) calculation. For DOS calculations, the tetrahedron method with Bl\"ochl corrections was employed~\cite{Bloechl.1994th}. The DFT$+U$ scheme~\cite{Anisimov.1991,Dudarev.1998} was adopted for the treatment of Mn $3d$ orbitals, with the parameter $U_\mathrm{eff} = U - J$ equal to $5$\,eV. The van der Waals interactions were incorporated by the semi-empirical approach of Grimme through the D2 correction~\cite{Grimme.2006}.  

The bulk structures were fully relaxed until forces became smaller than $0.01$\,eV\,\AA$^{-1}$. The convergence criteria for energy was set equal to $10^{-5}$\,eV. When studying bulk MPX$_3$, a $12 \times 12 \times 4$ k-mesh was used in the case of ionic relaxations and $24 \times 24 \times 8$ for single point calculations, respectively.

To model adsorption on top of 3D-MnPX$_3$ the supercells were used, which have a $(2\times2)$ surface periodicity and contain $7$ monolayers of MnPX$_3$. The lattice constant in the lateral plane was set according to the optimised lattice constants, $a(\mathrm{MnPS}_3)=6.075$\,\AA\ and $a(\mathrm{MnPSe}_3)=6.402$\,\AA\, respectively. A vacuum gap in all cases was at least $30$\,\AA. Li or F atoms were adsorbed from one side of the slab. During the structural optimisation procedure the coordinates of the adsorbate as well as  those of Mn, P, S/Se belonging to the surface layer were fully relaxed; the remaining ions were fixed at their bulk positions. The convergence criteria were identical to those used for bulk optimisation. A $6 \times 6 \times 1$ k-mesh was used in the case of ionic relaxations and $12 \times 12 \times 1$ for single point calculations, respectively.

The magnetic properties are characterised by an effective spin Hamiltonian on a hexagonal lattice~\cite{Sivadas.2015}:
\begin{equation} 
	\nonumber
	H= \sum_{\langle i,j \rangle} J_1\, \vec{S_i} \cdot \vec{S_j} + \sum_{\langle\langle i,j \rangle\rangle} J_2\, \vec{S_i} \cdot \vec{S_j} + \sum_{\langle\langle\langle i,j \rangle\rangle\rangle}J_3\, \vec{S_i} \cdot \vec{S_j},
\end{equation}		
where ($ \vec{S_i}$ ) is the total spin magnetic moment of the atomic site of $i$, $J_{i,j}$ are magnetic exchange coupling between two local spins. $J_1$, $J_2$, and $J_3$ represent the interacting with three nearest neighboring, six next-nearest neighboring, and three third-nearest neighboring Mn atoms, respectively. In order to get the magnetic exchange coupling parameters ($J_n$) the following equations were used~\cite{Yang.2020uc}:

\begin{equation}\nonumber
\begin{aligned}
&J_1 = -\frac{E_\mathrm{FM}-E_\mathrm{nAFM}+E_\mathrm{zAFM}-E_\mathrm{sAFM}}{16S^2}\,, \\[0.2cm]
&J_2 = -\frac{E_\mathrm{FM}+E_\mathrm{nAFM}-(E_\mathrm{zAFM}+E_\mathrm{sAFM})}{32S^2}\,, \\[0.2cm]
&J_3 = -\frac{E_\mathrm{FM}-E_\mathrm{nAFM}-3(E_\mathrm{zAFM}-E_\mathrm{sAFM})}{48S^2}\,. \\[0.2cm]
\end{aligned}
\end{equation}
Here: $E_\mathrm{FM}$, $E_\mathrm{nAFM}$, $E_\mathrm{zAFM}$, $E_\mathrm{sAFM}$ are the total energies of FM, N\'eel antiferromagnetic, zigzag antiferromagnetic, and stripy antiferromagnetic configurations, respectively. 

To estimate the critical temperature ($T_\mathrm{N}$ or $T_\mathrm{C}$), Monte-Carlo simulations were performed within the Metropolis algorithm with periodic boundary conditions~\cite{Metropolis.1953}. The three exchange parameters $J_1$, $J_2$ and $J_3$ were used in a $64 \times 64$ superlattice. Upon the heat capacity $C_v(T) = (\langle E^2\rangle-\langle E\rangle^2)/k_BT^2$ reaching the equilibrium state at a given temperature, the $T_\mathrm{N}$ or $T_\mathrm{C}$ value can be extracted from the peak of the specific heat profile.

Bader charge analysis was performed using a grid-based method as implemented in the Bader v1.04 code~\cite{Tang.2009}.

To recover an effective primitive cell band structure picture from the supercell band structure, the BandsUP software was employed~\cite{Medeiros.2014,Medeiros.2015}.

\begin{figure}
\includegraphics[width=0.48\textwidth]{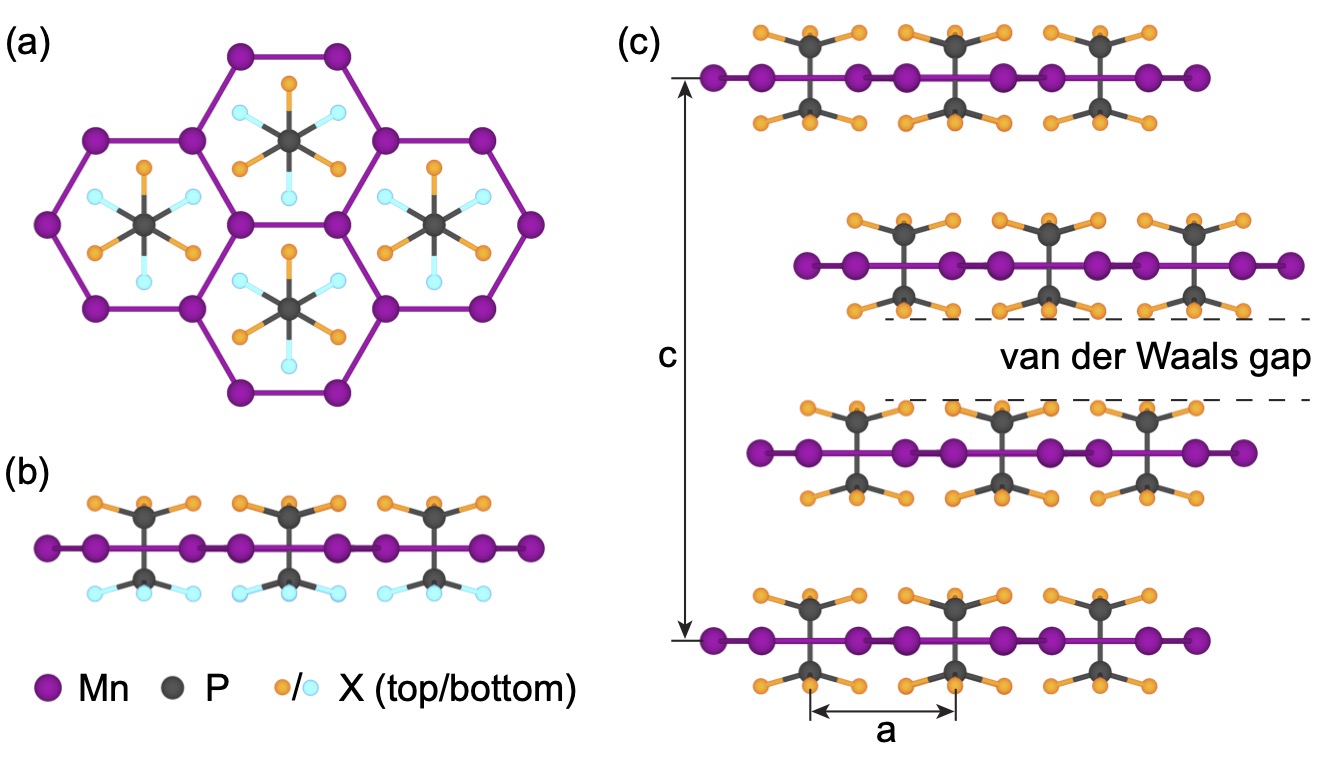}
\caption{\label{fig:MnPX3_2DandAdsPlaces} Lattice structure of MPX$_3$. (a) Top and  (b) side views of a single layer of MPX$_3$. (c) Crystal structure of 3D MPX$_3$. The in-plain and out-of-plane lattice constants are shown with letters $a$ and $c$, respectively. The van-der-Waals gap is indicated. Spheres of different size/colour represent ions of different type.}
\end{figure}

\section{Results}

The considered ``one-material''-based tunnel junction structure consists of several layers of MnPX$_3$ placed in the field-effect structure where electron/hole doping is induced in the left and right outermost layers of MnPX$_3$ using bias voltages applied to the conductive electrodes through the thin insulating (oxide) layers (Fig.~\ref{fig:scheme}a). As was previously discussed, the applied electric field breaks the spatial symmetry for the single TMT layer leading to the transformation from the AFM to HMF state~\cite{Li.2014}. In order to model the discussed effects of the electric-field induced doping and of the AFM/HMF transition for the left/right MnPX$_3$ layers, the adsorption of alkali and halogen atoms on the single and multilayer MnPX$_3$ is considered (Fig.~\ref{fig:scheme}b). In such a case, the adsorption of Li and F mimics the applied electric field and the HMF state in the 2D MnPX$_3$ layer is induced (Fig.~\ref{fig:scheme}c).

\begin{figure*}[ht]
\centering
\includegraphics[width=0.70\textwidth]{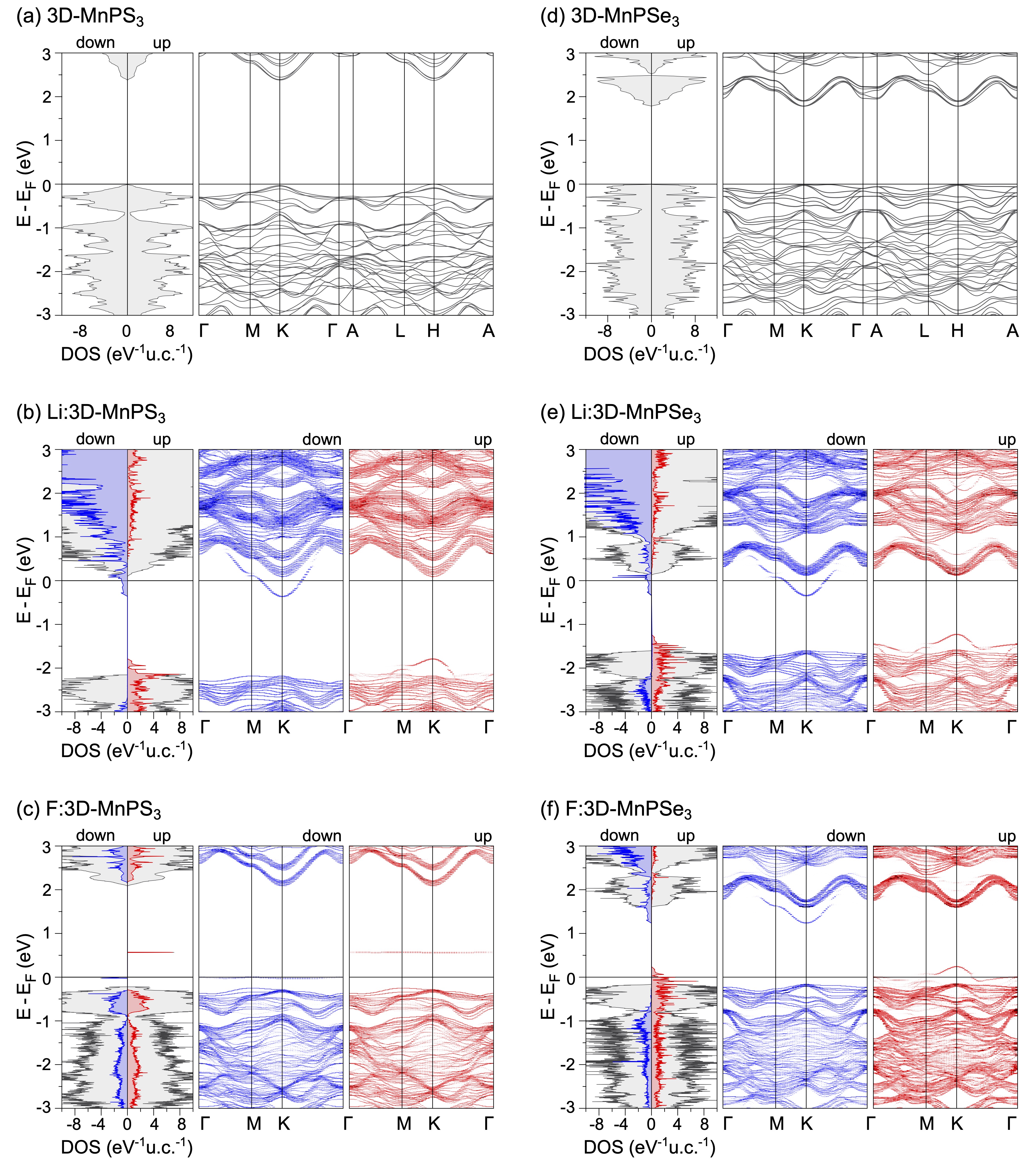}
\caption{\label{fig:MnPX3_3D_DOSandBANDS} Spin-resolved density of states and band structures obtained for the systems under consideration: (a) 3D-MnPS$_3$, (b) Li:3D-MnPS$_3$, (c) F:3D-MnPS$_3$, (d) 3D-MnPSe$_3$, (e) Li:3D-MnPSe$_3$, (f) F:3D-MnPSe$_3$. In DOS plots grey shaded areas indicate the total density of states and blue/red shaded areas indicate the spin-down/spin-up density of states projected onto atoms of the outermost MPX$_3$ layers.}
\end{figure*}

In the discussed structure, the 3D MnPX$_3$ bulk of $C2/m$ or $R\overline{3}$ symmetry for X\,=\,S or X\,=\,Se, respectively, consists of single 2D layers separated by the relatively large van der Waals gap of about $3.3$\,\AA\ between adjusting layers~\cite{Yang.2020uc} (Fig.~\ref{fig:MnPX3_2DandAdsPlaces} and Supplementary Fig.\,S2). In every single layer of MnPX$_3$, six Mn cations form a honeycomb lattice with a phosphorus dimer, which perpendicularly crosses the centre of every hexagon. The P -- P dimers are covalently bound to six sulphur (or selenium) atoms to form an ethane-like $(\mathrm{P}_2\mathrm{X}_6)^{4-}$ unit, where each P-atom is tetrahedrally coordinated with three S (or Se) atoms. Both bulk MnPS$_3$ and MnPSe$_3$ are the Ising-type AFM-N\'eel ordered direct wide-gap semiconductors with $E_g=2.37$\,eV/$T_N=87$\,K and $E_g=1.78$\,eV/$T_N=61$\,K ($E_g$ is a band gap), respectively (Fig.~\ref{fig:MnPX3_3D_DOSandBANDS}), in agreement with previous experimental and theoretical data~\cite{Du.2018,Joy.1992,Kim.2019gt,Liu.2020ajd}. The very weak interaction between single MnPX$_3$ layers conditioned by the large distance between layers and that every layer is terminated by the chalcogen atoms (S or Se) allows to conclude that layers are electronically and magnetically decoupled. In fact, the calculated energy differences between different inter-layer magnetic couplings, AFM and FM, are $0.5$\,meV and $0.27$\,meV for bulk MnPS$_3$ and MnPSe$_3$, respectively~\cite{Yang.2020uc}. (Further detailed information on the electronic and magnetic properties of bulk MnPX$_3$ (denoted as 3D-MnPX$_3$ further in the text) and MnPX$_3$ monolayers (denoted as 2D-MnPX$_3$ further in the text) can be found in Supplementary Information.)

\begin{figure}[ht]
\includegraphics[width=0.48\textwidth]{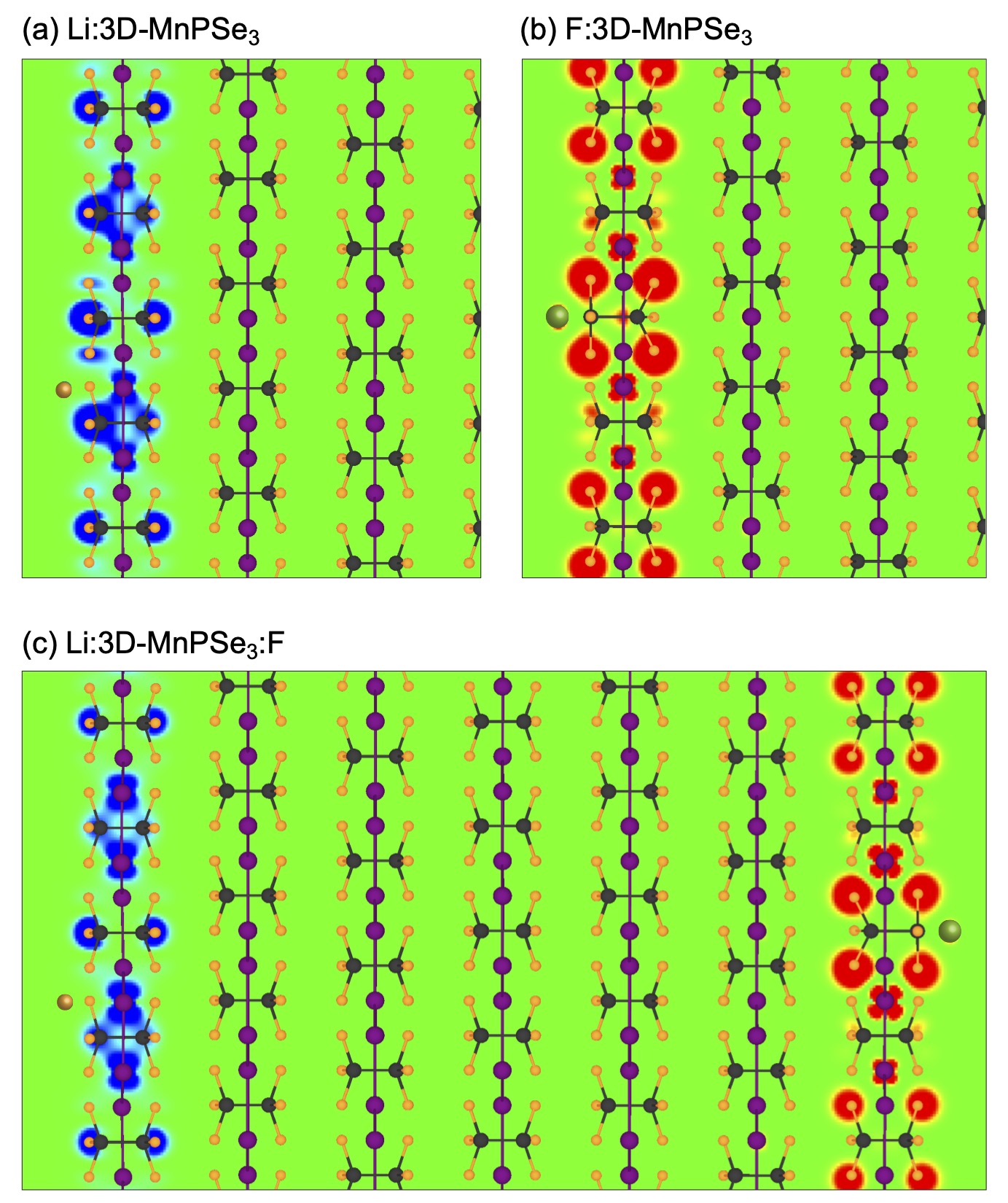}
\caption{\label{fig:MnPX3_3D_SpinDensityPlots} The real-space spin-density distribution maps overlaid with the respective structures for: (a) Li:3D-MnPSe$_3$, (b) F:3D-MnPSe$_3$, (c) Li:3D-MnPSe$_3$:F. The maps are colour coded as blue ($-0.001\,e/\mathrm{\AA}^3$), green ($0$), and red ($+0.001\,e/\mathrm{\AA}^3$).}
\end{figure}

Upon the adsorption of Li or F on 2D-MnPX$_3$ (in the $(2\times2)$ structure corresponding to the concentration of $1/4$\,ML), its magnetic and electronic structures are strongly modified. In the ground state structures, Li-atoms occupy the Mn adsorption site and F-atoms are coordinated directly above P atoms of the ethane-like $(\mathrm{P}_2\mathrm{X}_6)^{4-}$ unit of MnPX$_3$, respectively. In both adsorption systems, a significant charge transfer is found between adsorbate and 2D-MnPX$_3$, leading to the strong $n$- and $p$-doping in case of Li and F, respectively (Supplementary Figs.\,S4, S5, S7). For the both X\,=\,S/Se, adsorption of Li yields stabilisation of the FM state of MnPX$_3$ with a HMF behaviour in DOS and band structure -- the spin-down channel is metallic and the spin-up channel is semiconducting with $E_g= 2.27$\,eV and $1.34$\,eV for X\,=\,S and Se, respectively. In the case of the F-atoms adsorption, MnPS$_3$ remains semiconducting; however, the adsorption on MnPSe$_3$ converts the TMT monolayer in the HMF state with metallic spin-up channel and semiconducting spin-down channel with $E_g=1.95$\,eV. Monte-Carlo simulations give Curie temperature values of $T_\mathrm{C}=191\,\mathrm{K}/T_\mathrm{C}=113\,\mathrm{K}$ for Li:2D-MnPS$_3$/MnPSe$_3$ and $T_\mathrm{C}=96\,\mathrm{K}$ for F:2D-MnPSe$_3$, respectively (Fig.\,S6 of Supplementary Information). (The detailed information on the Li:2D-MnPX$_3$ and F:2D-MnPX$_3$ can be found in Supplementary Information.)

For the adsorption of Li and F on top of 3D-MnPX$_3$, the adsorbed atoms tend to occupy the same positions as for the considered cases of 2D-MnPX$_3$ -- Li-atoms are adsorbed above Mn adsorption site and F-atoms are adsorbed above P atoms. Therefore, as it follows from the above discussion on the weak inter-layer coupling in bulk MnPX$_3$, the otermost MnPX$_3$ layers which are in contact with adsorbates are stabilized in the FM state with the rest of the MnPX$_3$ stack remaining in the AFM semiconducting state. This situation is valid for both adsorbates on 3D-MnPX$_3$, as confirmed by the respective differences in total energy of $\Delta E = E_{FM} - E_{AFM} = -68$\,meV/u.c. ($-95$\,meV/u.c.) and $-124$\,meV/u.c. for Li:3D-MnPS$_3$ (Li:3D-MnPSe$_3$) and F:3D-MnPSe$_3$, respectively. The corresponding calculations and simulations also give the Curie temperature values of $T_\mathrm{C}=198\,\mathrm{K}/T_\mathrm{C}=126\,\mathrm{K}$ for Li:3D-MnPS$_3$/MnPSe$_3$ and $T_\mathrm{C}=85\,\mathrm{K}$ for F:3D-MnPSe$_3$, respectively.

The most intriguing is that the outermost MnPX$_3$ layers of the 3D stack, which are in contact with the Li or F adsorbates, demonstrate the HMF state with other deep layers in the insulating AFM state (Fig.~\ref{fig:MnPX3_3D_DOSandBANDS}). From these DOS and band structure plots one can see that for the upper TMT layer the spin-down (spin-up) channel is metallic and the opposite one is semiconducting for Li (F) adsorbates, respectively. The corresponding band gaps for semiconducting spin channels are $1.89$\,eV and $1.35$\,eV ($1.36$\,eV), for Li:3D-MnPS$_3$ and Li(F):3D-MnPSe$_3$, respectively (see also Supplementary Tab.\,S5). At the same time the band gap for the underlying layers is not less than $2.25$\,eV ($1.73$\,eV), which is very close to the one for bulk MnPS$_3$ (MnPSe$_3$). The results obtained for the system where two outermost MnPX$_3$ layers of the TMT stack are covered with Li and F (see Fig.~\ref{fig:scheme}b) are very similar to those obtained earlier -- these outermost layers demonstrate the HMF state with the sign of spin polarization depending on the kind of adsorbate and the rest of the bulk-like TMT stack remains in the insulating AFM state. The real-space spin-density distribution plots also confirm the main conclusions (Fig.~\ref{fig:MnPX3_3D_SpinDensityPlots}). Here we need to note that the HMF state in both cases of Li and F adsorption is achieved for the charge transfer of $5.5\times10^{-3}$ and $4.7\times10^{-3}$ carriers per atom (carrier concentration $1.5\times10^{13}\,\mathrm{cm}^{-2}$ and $1.3\times10^{13}\,\mathrm{cm}^{-2}$), respectively. Such carrier charge density modulations can be easily achieved in the experiment confirming the feasibility of realization of the electric-field driven ``one-material''-based magnetic tunnel junction using layered van der Waals wide band-gap antiferromagnetic MnPX$_3$.

\begin{figure}[ht]
\centering
\includegraphics[width=0.33\textwidth]{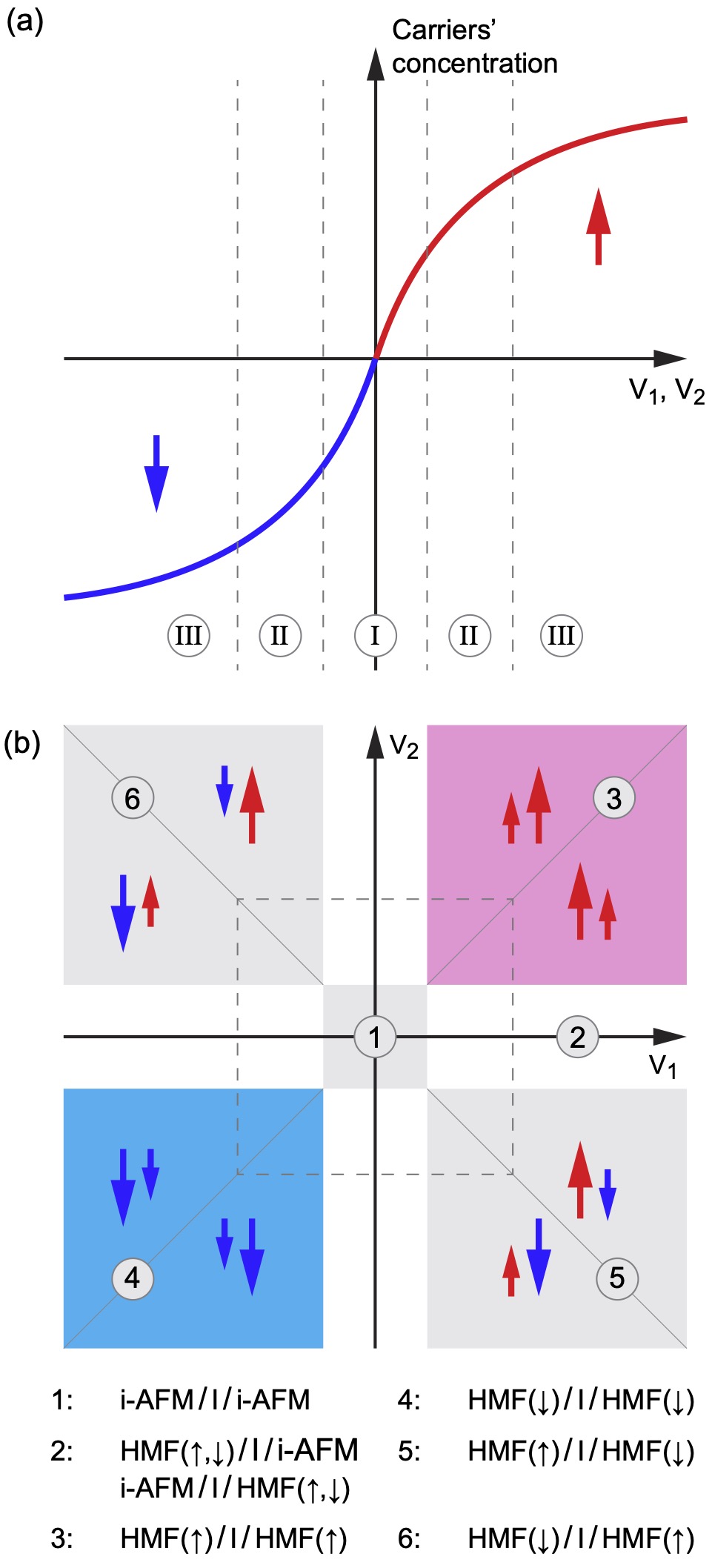}
\caption{\label{fig:MnPX3_3D_AdsPrincipleDiagram} State diagram for the double-sided biased MnPX$_3$ stack. (a) Schematic plot of carrier concentration as a function of bias voltage: (I) low bias voltage keeps all MnPX$_3$ stack in the i-AFM state, (II) HMF outermost MnPX$_3$ layer on i-AFM MnPX$_3$ stack, (III) HMF outermost MnPX$_3$ layer on PM or i-AFM MnPX$_3$ stack. (b) Depending on the sign of doping (sign of bias voltage) applied to the left/right outmost MnPX$_3$ layers, the different sign of spin polarization can be induced in the outmost layers. The direction (up/down) and the length of the arrows denote the sign of the spin polarization and the level of the carriers doping (see text for details).}
\end{figure}

\section{Discussion}

Following the analysis of the electronic structures of Li:3D-MnPX$_3$ and F:3D-MnPX$_3$, the functionality diagram for the sandwiched MnPX$_3$ stack can be drawn (Fig.~\ref{fig:MnPX3_3D_AdsPrincipleDiagram}). Depending on the sign of the bias voltage (and sign of doping, respectively) applied to the left/right outermost MnPX$_3$ layers, the different sign of spin polarization can be induced in these layers (Fig.~\ref{fig:MnPX3_3D_AdsPrincipleDiagram}a). Note, the outermost layers become HMF with a spin poarization of $\pm100$\% and the rest of the bulk remains insulating AFM (i-AFM). Using obtained theoretical results the different combinations of spin-polarization and doping levels allow to build very rich state diagram for the double-sided biased MnPX$_3$ stack (Fig.~\ref{fig:MnPX3_3D_AdsPrincipleDiagram}b). Here, the direction (up/down) and the length of the arrows denote the sign of the spin polarization (``$+/-$'') and the level of the carriers doping of the outmost MnPX$_3$ layers upon applied bias. For example, when level of the doping is low for both layers, then we simply have a complete insulating AFM stack of MnPX$_s$ (state 1). If one of the outermost layers is biased enough to be converted to the HMF state, but the opposite one is under-biased, then one obtains the HMF/i-AFM configuration, depending on the sign of spin polarization of the single outermost MnPX$_3$ layer (state 2). If two outermost MnPX$_3$ layers are biased to the HMF state with the same signs of spin polarization, then one of the states marked by 3 and 4 are realized. If outermost MnPX$_3$ layers are oppositely biased to the states when layers have different signs of spin polarization, then states marked by 5 and 6 are observed. From this consideration we can conclude that such electric-filed driven structure has TMR value of $100$\%. It is interesting to note that according to our findings, the Curie temperature of the biased (doped) MnPX$_3$ layers depends on the doping level -- the higher the doping the higher the $T_C$ value. Therefore, additionally to the previously considered cases of, e.\,g., HMF($\uparrow$)/i-AFM/HMF($\uparrow$) (state 3), the HMF($\uparrow$)/i-PM/HMF($\uparrow$) configuration (i-PM: insulating paramagnetic state of MnPX$_3$ above $T_N$) can be also realized at higher temperatures. The border between them is schematically shown by the dashed rectangle in Fig.~\ref{fig:MnPX3_3D_AdsPrincipleDiagram}b demarcating low and high levels of doping for the outermost MnPX$_3$ layers.

\section{Conclusions}

In summary, we propose the elegant approach for the realization of the electric-field driven ``one-material''-based magnetic tunnel junction. Here, several layers of the layered van der Waals TMT materials MnPX$_3$, which are insulating AFM, are used. Upon the applied bias voltage the outermost layers of the MnPX$_3$ stack can be converted from the insulating AFM state into the HMF and the sign of spin polarization depends on the direction of the applied electric field. Such combinations of bias voltages, and thus the spin polarization of the outer layers, give a rich states' diagram for the proposed ``one-material''-based magnetic tunnel junction, allowing to precisely control the state of this system. These results also offer an obvious potential for technological application as these systems possess a naturally occurring ideal HMF/insulating-AFM and HMF/insulating-PM interfaces. Such an ideal interfaces reduces unwanted scattering of conduction electrons in possible electronic applications.


\balance

\section*{Author contributions}
Y.D. proposed the study. E.V., Y.J., M.Y. performed the calculations. All authors discussed and analyzed the results and contributed to writing the manuscript.

\section*{Competing interests}
There are no conflicts to declare

\section*{Acknowledgments}

This work was supported by the National Natural Science Foundation of China (Grant No. 21973059). 
Y.D. and E.V. thank the "ENSEMBLE3 - Centre of Excellence for nanophotonics, advanced materials and novel crystal growth-based technologies" project (GA No. MAB/2020/14) carried out within the International Research Agendas programme of the Foundation for Polish Science co-financed by the European Union under the European Regional Development Fund and the European Union's Horizon 2020 research and innovation programme Teaming for Excellence (GA. No. 857543) for support of this work. 
The North-German Supercomputing Alliance (HLRN) is acknowledged for providing computer time.

\bibliographystyle{rsc} 

\providecommand*{\mcitethebibliography}{\thebibliography}
\csname @ifundefined\endcsname{endmcitethebibliography}
{\let\endmcitethebibliography\endthebibliography}{}

\end{document}